\documentclass[twocolumn,journal]{IEEEtran}

\usepackage{amsfonts,amsmath,amssymb,amsthm,algorithmic}
\usepackage{amsbsy}
\usepackage{tikz}
\usetikzlibrary{decorations.shapes,decorations.pathreplacing,decorations.markings}
\usetikzlibrary{intersections}
\usepackage{graphicx,multirow,bm}
\usepackage{color}
\usepackage[noadjust]{cite}

\makeatletter
\newtheoremstyle{mythm}{3pt}{3pt}{}{16pt}{\bfseries}{:}{.5em}{}
\theoremstyle{mythm}
\newtheorem{theorem}{Theorem}
\newtheorem{example}{Example}
\newtheorem{definition}{Definition}

\newtheorem{proposition}{Proposition}
\newtheorem{corollary}{Corollary}
\newtheorem{lemma}{Lemma}

\newtheorem{construction}{Construction}

\DeclareMathOperator{\diag}{diag}

\DeclareMathOperator{\norm}{Norm}

\DeclareMathOperator{\rank}{rank}

\DeclareMathOperator{\dec}{Dec}

\renewcommand{\leq}{\leqslant}

\renewcommand{\geq}{\geqslant}

\newcommand{\bC}{\bm{C}}

\newcommand{\cB}{\mathcal{B}}
\newcommand{\cC}{\mathcal{C}}
\newcommand{\cD}{\mathcal{D}}
\newcommand{\cE}{\mathcal{E}}

\newcommand{\cS}{\mathcal{S}}

\newcommand{\bbeta}{{\bm \beta}}

\newcommand{\N}{\mathbb{N}}

\newcommand{\F}{\mathbb{F}}
\newcommand{\parenv}[1]{\left( #1 \right)}

\newcommand{\ceilenv}[1]{\left\lceil #1 \right\rceil}
\newcommand{\floorenv}[1]{\left\lfloor #1 \right\rfloor}

\newcommand{\SR}{\mathrm{SR}}
\DeclareMathOperator{\wt}{wt}

\begin{document}

\title{A Construction of Maximally Recoverable Codes with Order-Optimal Field Size
\author{Han Cai, \IEEEmembership{Member,~IEEE}, Ying Miao, Moshe Schwartz, \IEEEmembership{Senior Member,~IEEE},\\ and Xiaohu Tang, \IEEEmembership{Senior Member,~IEEE}}
\thanks{H. Cai and X. Tang are with the School of Information Science and Technology,
  Southwest Jiaotong University, Chengdu, 610031, China (e-mail: hancai@aliyun.com; xhutang@swjtu.edu.cn).}
\thanks{M. Schwartz is with the School of Electrical and
Computer Engineering, Ben-Gurion University of the Negev, Beer Sheva
8410501, Israel (e-mail: schwartz@ee.bgu.ac.il).}
\thanks{Y. Miao is with the Faculty of Engineering, Information and Systems,
University of Tsukuba, Tennodai 1-1-1, Tsukuba 305-8573, Japan (e-mail: miao@sk.tsukuba.ac.jp).}
\thanks{This work was supported in part by a German Israeli Project Cooperation (DIP) grant under grant no.~PE2398/1-1,
a JSPS Grant-in-Aid
for Scientific Research (B) under Grant 18H01133, and a National Natural 
Science Foundation of China under Grant 61871331.}
}

\maketitle

\begin{abstract}
  We construct maximally recoverable codes (corresponding to partial
  MDS codes) which are based on linearized Reed-Solomon codes. The new
  codes have a smaller field size requirement compared with known
  constructions. For certain asymptotic regimes, the constructed codes
  have order-optimal alphabet size, asymptotically matching the known
  lower bound.
\end{abstract}

\section{Introduction}

\IEEEPARstart{D}{istributed} storage systems use erasure codes to
recover from node failures. Compared with the naive replication
solution, erasure-correcting codes, such as the maximum distance
separable (MDS) codes, can provide similar protection ability but with
a far smaller redundancy.  However, as the scale of system grows, new
challenges arise for MDS codes, such as repair bandwidth
\cite{sathiamoorthy2013xoring} and repair complexity
\cite{luby2019liquid}, due to the large number of nodes that need to
be contacted during the recovery process - even for a single erased
node.

One of the approaches that have been suggested to overcome those
issues is locally repairable codes
(LRCs)~\cite{gopalan2012locality}. In such a code, $k$ information
symbols are encoded into $n$ code symbols, which are arranged in
repair sets (perhaps overlapping) of size $r+\delta-1$.  Each repair
set is capable of recovering from $\delta-1$ erasures by using the
contents of the $r$ non-erased code symbols. Those codes are called
LRCs with $(r,\delta)$-locality.  Compared with MDS codes, even to
recover just one erasure, LRCs may
dramatically reduce the required repair bandwidth and repair
complexity, since for MDS codes we always need to contact $k$ code
symbols, whereas in LRCs we only
contact $r\ll k$ code symbols. For instances, in Microsoft Azure, an
LRC with $n=16$, $k=12$, $r=6$, and $\delta=2$, is used to reduce the
repair bandwidth~\cite{huang2012erasure}.

The original definition of LRCs with $(r,\delta=2)$-locality was
introduced in~\cite{gopalan2012locality}. Several generalizations have
followed later. The definition of LRCs was expanded to
$(r,\delta)$-locality with $\delta>2$ in \cite{prakash2012optimal}, to
allow repair sets to recover from more than one erasure. The concept
of availability was studied
in~\cite{wang2014repair,rawat2016locality,cai2019optimal} to allow
simultaneous recovery of a given code symbol from multiple repair
sets. To allow different requirements for local recovery, hierarchical
and unequal locality were introduced in~\cite{sasidharan2015codes}
and~\cite{zeh2016bounds,kim2018locally}, respectively. Over the past
decade, many bounds and constructions for LRCs have been introduced,
e.g.,
\cite{huang2013pyramid,rawat2013optimal,tamo2014family,cadambe2015bounds,wang2015integer,
li2019construction,xing2019construction,cai2020optimal,cai2021optimal_GPMDS,
chen2020improved,hao2020bounds,neri2020random}
for $(r,\delta)$-locality
\cite{rawat2016locality,TaBaFr16bounds,cai2018optimal,cai2019optimal,jin2019construction}
for multiple repair sets,
\cite{sasidharan2015codes,luo2020optimal,zhang2020constructions,chen2020cyclic}
for hierarchical locality, and~\cite{zeh2016bounds,kim2018locally} for
unequal locality.

As is usually the case, locality comes at a cost of reduced code rate
and minimum Hamming distance. It was shown
in~\cite{gopalan2012locality} that, except for trivial cases, the
minimum Hamming distance of LRCs cannot attain the well known
Singleton bound~\cite{singleton1964maximum}. To make the most out of
this restriction, one natural problem is whether LRCs can recover from
some predetermined erasure patterns beyond those guaranteed by their
minimum Hamming distance.  A subclass of LRCs named \emph{maximally
  recoverable} (MR) codes~\cite{gopalan2012locality} offer a positive
answer to this question, by correcting the maximal possible set of
erasure patterns beyond the minimum Hamming distance. \emph{Partial
  MDS} (PMDS) codes~\cite{blaum2013partial}, that form a subclass of
MR codes, improve the storage efficiency of RAID systems, where $h$
extra erasures may be recovered in addition to $\delta-1$ erasures in
each repair set.

Motivated by their efficiency and applicability, $[n,k,d]_q$ MR codes
with $(r,\delta)$-locality, and $h$ global parity-check symbols, have
received much attention over the recent few years, { where $[n,k,d]_q$ denotes a linear
code with length $n$, dimension $k$, and minimum Hamming
distance $d$, over a field of size $q$.}  For $[n,k,d]_q$ MR codes
with $(r,\delta)$-locality, of particular interest have been the asymptotic regime in which $h$ and $\delta$ are
constants, and the goal to construct codes with the smallest possible
field size { $q$}. { For the case of $h=1$, MR
codes were constructed over a finite field of size
$q=\Theta(r+\delta-1)$~\cite{blaum2013partial} and a characterization was
given in \cite{horlemann2020complete}.}   When $h=2$, MR codes were constructed
in~\cite{blaum2016construction} with $q=\Theta(n(\delta-1))$, and
later, with $q=\Theta(n)$~\cite{gopi2020maximally}
(see~\cite{hu2016new} for $n=2(r+\delta-1)$).  For $h=3$, MR codes
were constructed with $q=\Theta(n^{3/2})$ for a constant
$r+\delta-1$, and  $q=\Theta(n^3)$ for an odd $q$
\cite{gopi2020maximally}. For the case of $\delta=2$, constructions
for MR codes were provided for finite fields with
size $q=\Theta(k^{h-1})$~\cite{gopalan2014explicit}.
For the case $r=2$, the existence of MR codes
was proved in~\cite{bogart2020constructing} using a field of size $q=\Theta(n^{h-1})$.
For general
$\delta$ and $h$, a construction of MR codes with flexible parameters
was introduced based on Gabidulin codes~\cite{calis2016general}, which
requires a field with size $q=\Theta((r+\delta-1)^{nr/(r+\delta-1)})$.
Additionally, MR codes were constructed over finite fields with size
$q=\Theta((r+\delta-1)n^{h\delta-1})$ and
$q=\Theta(\max(\frac{n}{r+\delta-1},(r+\delta-1)^{h+\delta-1})^h)$~\cite{gabrys2018constructions}. In~\cite{guruswami2020constructions},
MR codes were constructed with
$q=\Theta(\max(\frac{n}{r+\delta-1},(2r)^{h+\delta-1})^{\min(\frac{n}{r+\delta-1},h)})$
and
$q=\Theta(\max(\frac{n}{r+\delta-1},(2r)^{r+\delta-1})^{\min(\frac{n}{r+\delta-1},h)})$,
respectively.  Recently, based on linearized Reed-Solomon codes, MR
codes were constructed with
$q=\Theta(\max(r+\delta-1,\frac{n}{r+\delta-1})^r)$~\cite{martinez2019universal},
which is independent of the number of global parity-check symbols $h$,
thus outperforming other known constructions when $h$ is relatively
large, namely, $h\geq r$. In \cite{holzbaur2021partial}, the authors construct MR codes with
optimal repairing bandwidth inside repair sets. The parameters
of MR codes from the known constructions, as well as a new one of this paper, are listed in Table~\ref{tab:comp}.

\begin{table*}[t]
  \caption{Known $(n,r,h,\delta,q)$-MR codes (PMDS codes) in the
    asymptotic regime where $h$ and $\delta$ are constant, and where $m\triangleq\frac{n}{r+\delta-1}$}
  \label{tab:comp}
  \begin{center}
    \begin{tabular}{|c|c|c|c|c|c|c|}
      \hline
      \multirow{2}{*}{$r$} & \multirow{2}{*}{$\delta$} & \multirow{2}{*}{$h$} & \multirow{2}{*}{Size of Alphabet ($q$)}
      & Cases with order & \multirow{2}{*}{Restrictions} &\multirow{2}{*}{Ref}. \\
      &&&&optimal field size&&\\
      \hline\hline
      any & any & $1$& $\Theta(r+\delta-1)$  & {all possible cases} & & \cite[Thm. 5.4]{blaum2013partial} \\
      \hline
      \multirow{4}{*}{any} & \multirow{4}{*}{any} & \multirow{4}{*}{$2$}& $\Theta(n\delta)$  & all possible cases &  & \cite[Thm. 7]{blaum2016construction} \\
      \cline{4-7}
      & & & $\Theta(n)$ &
      {all possible cases} & $q$ is odd& \cite[Thm. IV.4]{gopi2020maximally} \\
      \cline{4-7}
            &  & & $n\cdot\exp(O(\sqrt{\log n}))$ &
      {None} & $q$ is even& \cite[Thm. IV.4]{gopi2020maximally} \\
      \cline{4-7}
       & & & $\Theta(n)$ &
      all possible cases & $q$ is even, $n=\Theta(m^2)$& Construction~\ref{cons_PMDS}\\
      \hline
      \multirow{4}{*}{any} & \multirow{4}{*}{any} & \multirow{4}{*}{$3$}& $\Theta(n^{3/2})$  &
      \multirow{4}{*}{None} & $r$ is a constant, $q$ is even & \cite[Cor. 23]{gopalan2014explicit} \\
      \cline{4-4}
      \cline{6-7}
      & & &$\Theta(n^3)$ &  & $q$ is odd & \cite[Thm. V.4]{gopi2020maximally} \\
      \cline{4-4}
      \cline{6-7}
       & & &$n^3\cdot\exp(O(\sqrt{\log n}))$ &  & $q$ is even & \cite[Thm. V.4]{gopi2020maximally} \\
       \cline{4-4}
      \cline{6-7}
      & & &$\Theta(n^3)$ &  &   & Construction~\ref{cons_PMDS} \\
      \hline
       2& any& any& $O(n^{h-1})$& $h=2$& $m\geq h$ &  \cite[Cor. 7.14]{bogart2020constructing}\\
      \hline
      \multirow{2}{*}{any} & \multirow{2}{*}{$2$} & \multirow{2}{*}{any}
      & \multirow{2}{*}{$\Theta(k^{\lceil (h-1)(1-1/2^r)\rceil})$}  & $h=3$,  &  & \multirow{2}{*}{\cite[Cor. 18]{gopalan2014explicit}} \\
      & & & & $m\geq 3$ is a constant &  &\\
      \hline
      any & any & any& $\Theta((r+\delta-1)^{nr/(r+\delta-1)})$  & None &  & \cite[Cor. 11]{calis2016general} \\
      \hline
      any & any & any& $\Theta((r+\delta-1)n^{h\delta-1})$  & None &$q_1=r+\delta-1$, $2n=q_1^t$  & \cite[Lem. 7]{gabrys2018constructions}\\
      \hline
      any & any & any& $\Theta(\max\{m,(r+\delta-1)^{h+\delta-1}\}^h)$  & None &$q_1=r+\delta-1$, $m+1=q^t_1$  & \cite[Cor. 10]{gabrys2018constructions}\\
      \hline
      any & any & any& $\Theta(\max\{m,(2(r+\delta-1))^{h+\delta-1}\}^{\min(m,h)})$  & None &  & \cite[Thm. 17]{guruswami2020constructions}\\
      \hline
      any & any & any& $\Theta(\max\{m,(2(r+\delta-1))^{r+\delta-1}\}^{\min(m,h)})$  & None &  & \cite[Thm. 19]{guruswami2020constructions}\\
      \hline
      any & any & any& $\Theta(\max\{r+\delta-1,m\}^r)$  & None &  & \cite[Cor. 8]{martinez2019universal}\\
      \hline
      \multirow{2}{*}{any} & \multirow{2}{*}{any} & \multirow{2}{*}{any}&
      \multirow{2}{*}{$\Theta(\max\{r+\delta-1,m\}^h)$}
      & $h\leq \min\{m,\delta+1\}$,&& \multirow{2}{*}{Construction~\ref{cons_PMDS}}\\
      &&&& $n=\Theta(m^2)$ &&\\
      \hline
    \end{tabular}
  \end{center}
\end{table*}

However, there is still an asymptotic gap between the known lower
bounds on the minimum field size of MR codes~\cite{gopi2020maximally}
and the known constructions. The main contribution of this paper is a
new construction of MR codes over small finite fields when $h$ is
relatively small, namely, $h<r$. Our construction is inspired by the
construction in~\cite{martinez2019universal}, and we also use
linearized Reed-Solomon codes, yielding MR codes with field size
$\Theta(\max\{r+\delta-1,\frac{n}{r+\delta-1}\}^{h})$.  Compared with the
known constructions
in~\cite{calis2016general,gabrys2018constructions,guruswami2020constructions,martinez2019universal},
our construction generates MR codes with a smaller field size. In
particular, our MR codes have order-optimal field size, asymptotically
matching the lower bound in~\cite{gopi2020maximally} when
$r+\delta-1=\Theta(\sqrt{n})$ and
$h\leq\min\{\frac{n}{r+\delta-1},\delta+1\}$. Our construction also
answers an open problem from~\cite{gopi2020maximally}, by
providing MR codes over a field with even (or odd) characteristic.
We would like to comment that shortly after we published our results, we learned that~\cite{gopi} have independently obtained a similar construction.

The remainder of this paper is organized as follows. Section
\ref{sec-preliminaries} introduces basic notation and definitions of
LRCs and MR codes, known bounds, as well as required facts on
linearized Reed-Solomon codes.  Section~\ref{sec-construction-MRC}
presents our construction of MR codes.  Section \ref{sec-conclusion}
concludes this paper by summarizing and comparing our codes with the
known codes, and discussing important cases.

\section{Preliminaries}
\label{sec-preliminaries}

Let us introduce the notation, definitions, and known results used
throughout this paper. For a positive integer $n$, we denote
$[n]\triangleq\{1,2,\cdots,n\}$. If $q$ is a prime power, let $\F_q$
denote the finite field with $q$ elements.

An $[n,k]_q$ linear code $\cC$ over $\F_q$ is a $k$-dimensional
subspace of $\F_q^n$ with a $k\times n$ generator matrix
$G=(\mathbf{g}_1,\mathbf{g}_2,\cdots,\mathbf{g}_{n})$, where
$\mathbf{g}_i$ is a column vector of length $k$ for all
$i\in[n]$. Specifically, $\cC$ is called an $[n,k,d]_q$ linear code if
the minimum Hamming distance of $\cC$ is $d$. { For an $m\times n$ matrix $A=(A_1, A_2, \dots, A_n)\in \F^{m\times n}_q$ and
$I\subseteq [n]$, let $A|_{I}$ denote the projection of $A$ upon columns specified by $I$,
i.e., $A|_{I}=(A_{i})_{i\in I}$.}
{  For any
codeword $C=(c_1,c_2,\ldots,c_n)\in \cC$, we say that $c_i$, $i\in[n]$, is
the $i$th code symbol.
}

\begin{definition}[\cite{gopalan2012locality,prakash2012optimal}]
  The $i$th code symbol of an $[n, k, d]_q$ linear code $\cC$ is said
  to have $(r, \delta)$-\emph{locality} if there exists a subset
  $S_i\subseteq [n]$ (an $(r,\delta)$-\emph{repair set}) such that
  \begin{itemize}
  \item $i\in S_i$ and $|S_i|\leq r+\delta-1$; and
  \item The minimum Hamming distance of the punctured code
    $\cC|_{S_i}$ obtained by deleting the code symbols $c_j$ ($j \in
    [n]\setminus S_i$) is at least $\delta$.
  \end{itemize}
  Furthermore, an $[n,k,d]_q$ linear code $\cC$ is said to have
  information $(r,\delta)$-locality (denoted as
  $(r,\delta)_i$-locality) if there exists a $k$-subset $I\subseteq
  [n]$ with $ \rank(G|_I)=k$ such that for each $i\in I$, the $i$th code
  symbol has $(r, \delta)$-locality, and all symbol
  $(r,\delta)$-locality (denoted as $(r,\delta)_a$-locality) if all
  the $n$ code symbols have $(r,\delta)$-locality.
\end{definition}

An upper bound on the minimum Hamming distance of linear codes with
$(r,\delta)_i$-locality was derived as follows (for $\delta=2$ in
\cite{gopalan2012locality}, and for general $\delta$ in
\cite{prakash2012optimal}):
\begin{lemma}[\cite{gopalan2012locality,prakash2012optimal}] \label{lemma_bound_i}
  For an $[n,k,d]_q$ code $\cC$ with $(r,\delta)_i$-locality,
  \begin{equation}\label{eqn_bound_all_local}
    d\leq n-k+1-\left(\left\lceil\frac{k}{r}\right\rceil-1\right)(\delta-1).
  \end{equation}
\end{lemma}

A linear code with information $(r,\delta)_i$-locality (or
$(r,\delta)_a$-locality) is said to be optimal if its minimum Hamming
distance achieves the bound in~\eqref{eqn_bound_all_local}.

\begin{definition}
  \label{def:mrcodes}
  Let $\cC$ be an $[n,k,d]_q$ code with $(r,\delta)_a$-locality, and
  define $\cS\triangleq\{S_i ~:~ i\in [n]\}$, where $S_i$ is an
  $(r,\delta)$-repair set for coordinate $i$. The code $\cC$ is said
  to be a \emph{maximally recoverable} (MR) code if $\cS$ is a
  partition of $[n]$, and for any $R_i\subseteq S_i$ such that
  $|S_i\setminus R_i|=\delta-1$, the punctured code $\cC|_{\cup_{1\leq
      i\leq n}R_i}$ is an MDS code.
\end{definition}

Of particular interest are MR codes for which $\cS$ is a partition of
$[n]$ with equal-size parts.

\begin{definition}
  Let $\cC$ be an $[n,k,d]_q$ MR code, as in
  Definition~\ref{def:mrcodes}.  If each $S_i\in\cS$ is of size
  $|S_i|=r+\delta-1$, then $r+\delta-1|n$. Define
  \begin{eqnarray*}
    m\triangleq \frac{n}{r+\delta-1}, \qquad h\triangleq mr-k.
  \end{eqnarray*}
  Then $\cC$ is said to be an $(n,r,h,\delta,q)$-MR code.
\end{definition}

{  We note that in general, MR codes need not have repair sets of equal
size, nor do the repair sets have to form a partition of $[n]$. In
this paper we choose to follow the more restrictive definition
from~\cite{gopalan2012locality,gopalan2014explicit}.}

We also note that it is easy to verify that
$(n,r,h,\delta,q)$-MR codes are optimal $[n,k,d]_q$ LRCs with
$(r,\delta)_a$-locality.
We can regard each codeword of an $(n,r,h,\delta,q)$-MR code, as an
$m\times(r+\delta-1)$ array, by placing each repair set in $\cS$ as a
row. When viewed in this way, $(n,r,h,\delta,q)$-MR codes match the
definition of partial MDS (PMDS) codes, as defined in
\cite{blaum2013partial}, where in a codeword, each entry of the array
corresponds to a sector, and each column of the array corresponds to a
disk.


For the sake of completeness, we would like to mention that aside from
PMDS codes, there are other codes with locality that can recover from
predetermined erasure patterns beyond the minimum Hamming distances
\cite{plank2014sector,li2014stair,gopalan2017maximally,cai2021optimal_GPMDS}. As
an example, sector-disk (SD) codes~\cite{plank2014sector} with
$(r,\delta)_a$-locality can correct $\delta-1$ disk erasures together
with any additional $h$ sector erasures, where $h$ denotes the number
of global parity-check symbols.

One interesting problem arising from the definition of MR codes is to
determine the minimum alphabet size for fixed $n$, $r$, $h$, and
$\delta$.  For the case $h=1$, it is easy to check that an
$(n,r,1,\delta,q)$-MR code is an optimal LRC with
$(r,\delta)_a$-locality and $d=\delta+1$, where $(r+\delta-1)|n$ and
$k=\frac{rn}{r+\delta-1}-1$. For this case, the field size requirement
may be as small as $q=\Theta(r+\delta-1)$, which is asymptotically
optimal for the simple reason that the punctured code over any repair
set together with the only global parity check is an
$[r+\delta,r,\delta+1]_q$ MDS code when $(r+\delta-1)|n$.  For the
case $h\geq 2$, in \cite{gopi2020maximally}, the following asymptotic
lower bounds on the field size are derived. We emphasize that here,
and throughout the paper, we assume $h$ and $\delta$ are constants.

\begin{lemma}[{\cite[Theorem I.1]{gopi2020maximally}}]
  \label{lemma_lower_bound_F}
  Let $h\geq 2$ and $\cC$ be an $(n,r,h,\delta,q)$-MR code. If
  $m\triangleq\frac{n}{r+\delta-1}\geq 2$, then
  \begin{equation*}
    q= \Omega(n r^{\varepsilon}),
  \end{equation*}
  where
  $\varepsilon=\min\{\delta-1,h-2\lceil\frac{h}{m}\rceil\}/\lceil\frac{h}{m}\rceil$,
  and $h$ and $\delta$ are regarded as constants. The above lower
  bound may be simplified as
  \begin{enumerate}
  \item If $m\geq h$:
    \[
    q= \Omega\parenv{n r^{\min\{\delta-1,h-2\}}}.
    \]
  \item If $m\leq h$, $m|h$, and $\delta-1\leq h-\frac{2h}{m}$:
    \[q= \Omega\parenv{n^{1+\frac{m(\delta-1)}{h}}}.\]
  \item If $m\leq h$, $m|h$, and $\delta-1> h-\frac{2h}{m}$:
    \[q= \Omega\parenv{n^{m-1}}.\]
  \end{enumerate}
\end{lemma}

\begin{definition}\label{def_ord_opt}
An $(n,r,h,\delta,q)$-MR code is \textit{order-optimal} if it attains
one of the bounds of Lemma \ref{lemma_lower_bound_F} asymptotically
for $h\geq 2$, or if it has $q=\Theta(r+\delta-1)$ for $h=1$.
\end{definition}

\subsection{ The Sum-Rank Metric and Linearized Reed-Solomon Codes}

We turn to introduce some necessary definitions for linearized
Reed-Solomon codes, which form the main tool used in this paper.
We first recall the definition of the sum-rank metric as
  defined in~\cite{nobrega2010multishot} and ~\cite{martinez2018skew}.

\begin{definition}[\cite{martinez2018skew}]
Let $\F_q$ be a subfield of $\F_{q_1}$ and $N$, $L_i$ for $1\leq i\leq
g$, be positive integers with $N=\sum_{i=1}^{g}L_i$.  Let
$\bC=(\bC_1,\bC_2,\ldots,\bC_{g})\in \F^{N}_{q_1}$, where $\bC_i\in
\F^{L_i}_{q_1}$ for $1\leq i\leq g$. The \emph{sum-rank weight} in
$\F^{N}_{q_1}$, with length partition $(L_1,L_2,\dots,L_{g})$, is
defined as
\begin{equation*}
\wt_{\SR}(\bC)=\sum_{i=1}^{g} \rank_{q}(\bC_{i}),
\end{equation*}
where $\rank_{q}(\bC_{i})$ denotes the rank of $\bC_i\in
\F^{L_i}_{q_1}$ over $\F_q$.  Furthermore, for $\bC,\bC'\in
\F^{N}_{q_1}$, define the \emph{sum-rank distance} as
\begin{equation*}
d_{\SR}(\bC,\bC')=\wt_{\SR}(\bC-\bC').
\end{equation*}
\end{definition}

For a code $\cC\subseteq \F^{N}_{q_1}$, with length partition
$(L_1,L_2,\ldots,L_g)$ as before, we define the minimum sum-rank
distance by
\begin{equation*}
d_{\SR}(\cC)=\min\left\{d_{\SR}(\bC,\bC')~:~\bC,\bC'\in\cC,\, \bC\ne \bC'\right\}.
\end{equation*}
In an analogy with the Hamming metric, there is also a Singleton bound
for the sum-rank metric codes.

\begin{lemma}[\cite{martinez2018skew}]
Let $q_1=q^m$ and $\cC\subseteq \F^{N}_{q_1}$. Then we have
\begin{equation*}
|\cC|\leq q^{m(N-d_{\SR}(\cC)+1)}.
\end{equation*}
\end{lemma}

Similar to MDS codes, codes that attain the above
Singleton bound with equality are called \emph{maximum sum-rank
  distance (MSRD) codes}~\cite{martinez2018skew}.

This general definition of the sum-rank metric includes the Hamming
metric as a special case when the length partition is $g=N$ and
$L_1=L_2=\dots=L_n=1$.  It also includes the rank metric as a special
case when the length partition is $g=1$ and $L_1=N$. In what follows,
we introduce one class of MSRD codes called \emph{linearized Reed-Solomon
codes} \cite{martinez2018skew}.

Let $\F_q\subseteq \F_{q_1}$ and define $\sigma:\F_{q_1}\rightarrow
\F_{q_1}$ as
\[\sigma(\alpha)\triangleq \alpha^q.\]
For any $\alpha\in \F_{q_1}$ and $i\in \N$, define
\[\norm_i(\alpha)\triangleq \sigma^{i-1}(\alpha)\cdots\sigma(\alpha)\alpha.\]
The $\F_q$-linear operator $\cD^i_\alpha:\F_{q_1}\rightarrow \F_{q_1}$
is defined by
\[\cD^{i}_\alpha(\beta)\triangleq\sigma^i(\beta)\norm_i(\alpha).\]
Let $\alpha\in\F_{q_1}$, and let
$\cB=(\beta_1,\beta_2,\cdots,\beta_L)\in \F^{L}_{q_1}$.  For $i \in \N
\cup \{0\}$ and $k, \ell\in \N$, where $\ell\leq L$, define the
matrices
\begin{equation}\label{Eqn_Def_D}
\begin{split}
&D(\alpha^i,\cB,k,\ell)\\
\triangleq&
\begin{pmatrix}
\beta_1 &\beta_2&\cdots&\beta_{\ell}\\
\cD^1_{\alpha^{i}}(\beta_1) &\cD^1_{\alpha^{i}}(\beta_2)& \cdots &\cD^1_{\alpha^{i}}(\beta_{\ell})\\
\vdots &\vdots& &\vdots\\
\cD^{k-1}_{\alpha^{i}}(\beta_1) &\cD^{k-1}_{\alpha^{i}}(\beta_2)& \cdots &\cD^{k-1}_{\alpha^{i}}(\beta_{\ell})\\
\end{pmatrix}{ \in \F^{k\times\ell}_{q_1}}.
\end{split}
\end{equation}

{ The matrix defined by~\eqref{Eqn_Def_D} satisfies the following column linearity:

  \begin{proposition}\label{prop_linearity}
    With the setting as in~\eqref{Eqn_Def_D}, for any $A\in \F^{\ell\times
      \ell_1}_{q}$ we have
\begin{equation*}
D(\alpha^i,\cB,k,\ell)A=D(\alpha^i,\cB|_{[\ell]} A,k,\ell_1).
\end{equation*}
\end{proposition}
\begin{IEEEproof}
Write $\cB|_{[\ell]} A=(\beta'_1,\beta'_2,\ldots, \beta'_{\ell_1})$. Then, by~\eqref{Eqn_Def_D},
\begin{equation*}
\begin{split}
&D(\alpha^i,\cB,k,\ell)A\\
=&\begin{pmatrix}
\beta_1 &\beta_2&\cdots&\beta_{\ell}\\
\cD^1_{\alpha^{i}}(\beta_1) &\cD^1_{\alpha^{i}}(\beta_2)& \cdots &\cD^1_{\alpha^{i}}(\beta_{\ell})\\
\vdots &\vdots& &\vdots\\
\cD^{k-1}_{\alpha^{i}}(\beta_1) &\cD^{k-1}_{\alpha^{i}}(\beta_2)& \cdots &\cD^{k-1}_{\alpha^{i}}(\beta_{\ell})\\
\end{pmatrix}A\\
&=\begin{pmatrix}
\beta'_1 &\beta'_2&\cdots&\beta'_{\ell_1}\\
\cD^1_{\alpha^{i}}(\beta'_1) &\cD^1_{\alpha^{i}}(\beta'_2)& \cdots &\cD^1_{\alpha^{i}}(\beta'_{\ell_1})\\
\vdots &\vdots& &\vdots\\
\cD^{k-1}_{\alpha^{i}}(\beta'_1) &\cD^{k-1}_{\alpha^{i}}(\beta'_2)& \cdots &\cD^{k-1}_{\alpha^{i}}(\beta'_{\ell_1})\\
\end{pmatrix}\\
&=D(\alpha^i,\cB|_{[\ell]} A,k,\ell_1).\\
\end{split}
\end{equation*}
\end{IEEEproof}}

\begin{definition}[\cite{martinez2018skew}]\label{def_LRS}
  For positive integers $N$, $M$, $L$, and $g$, let $N=L_1+L_2+ \cdots
  +L_g$, $g\leq q-1$, and $1\leq L_i\leq L\leq M$.  Set
  $\F_{q_1}=\F_{q^M}$. Let $\cB$ be a sequence of elements that are
  linearly independent over $\F_q$. Then the \emph{linearized
    Reed-Solomon code} with dimension $k$, primitive element
  $\gamma\in \F_{q^M}$, and basis $\cB$, is the linear code
  $\cC^{\sigma}_{L,k}(\cB,\gamma)\subseteq \F^N_{q^M}$ with generator
  matrix
  \begin{equation*}
  \begin{split}
  D=&\left(D(\gamma^0,\cB,k,L_1), D(\gamma^1,\cB,k,L_2),\right.\\
   &\qquad\left.\cdots, D(\gamma^{g-1},\cB,k,L_g)\right)_{k\times N}.
  \end{split}
  \end{equation*}
\end{definition}

We comment that Definition~\ref{def_LRS} is a \emph{narrow-sense}
linearized Reed-Solomon code, which suffices for this paper. For a
more general definition of linearized Reed-Solomon code the reader is
referred to~\cite{martinez2018skew}. We also point out that linearized
Reed-Solomon codes are MSRD codes \cite{martinez2018skew}. For more
details on sum-rank metric codes and their applications to LRCs, the
reader may refer to~\cite{martinez2019universal}.

Let $\diag(W_1,W_2,\cdots,W_g)$ denote the block-diagonal
matrix, whose main-diagonal blocks are $W_1,W_2,\cdots, W_g$, i.e.,
\begin{equation*}
\diag(W_1,W_2,\cdots,W_g)=
\parenv{\begin{matrix}
W_1 &0 &\cdots &0\\
0&W_2&\cdots &0\\
\vdots&\vdots&\ddots&\vdots\\
0&0&\cdots&W_g\\
\end{matrix}}.
\end{equation*}
{  Since linearized Reed-Solomon codes are MSRD codes, the
  dimension $k$ of the code $\cC$ is $k=N-d_{\SR}(\cC)+1$. When it
  comes to correcting erasures, if the non-erased part has sum-rank
  weight at least $k$, the code can correctly recover the
  codeword. This is more formally described in the following lemma
  from~\cite{martinez2019universal}.}

\begin{lemma}[\cite{martinez2019universal}]\label{theorem_LRS}
  Let $g\leq q-1$, and let $\cC^{\sigma}_{L,k}(\cB,\gamma)$ be the
  $[N,k,N-k+1]_{q^M}$ linearized Reed-Solomon code from
  Definition~\ref{def_LRS}, with $N=L_1+L_2+\cdots+L_g$, and $1\leq
  L_i\leq L\leq M$. Then for all integers $n_i\geq 1$, and all
  matrices $W_i\in \F_q^{L_i\times n_i}$, $i\in[g]$, satisfying
  \[
  \sum_{i=1}^{g}\rank(W_i)\geq k,
  \]
  there exists a decoder
  \[\dec:\cC^{\sigma}_{L,k}(\cB,\gamma)\diag(W_1,W_2,\cdots,W_g)\rightarrow \cC^{\sigma}_{L,k}(\cB,\gamma)\]
  such that
  \begin{equation*}
    \dec(C\diag(W_1,W_2,\cdots,W_g))=C \quad \text{for any }C\in \cC^{\sigma}_{L,k}(\cB,\gamma),
  \end{equation*}
  where
  \begin{equation*}
  \begin{split}
    &\cC^{\sigma}_{L,k}(\cB,\gamma)\diag(W_1,W_2,\cdots,W_g)\\
    \triangleq& \{C\diag(W_1,W_2,\cdots,W_g) ~:~ C\in \cC^{\sigma}_{L,k}(\cB,\gamma)\}.
  \end{split}
  \end{equation*}
\end{lemma}

{ Furthermore, when we analyze the case in which the
  non-erased part has sum rank less than $k$, we arrive at the following
  property of generator matrices for linearized Reed-Solomon codes,
  which is a direct application of the previous lemma.}

\begin{theorem}\label{thm_rank}
  Let $g\leq q-1$, and $D$ be  generator matrix of a linearized Reed-Solomon code from Definition~\ref{def_LRS}  with $N=L_1+L_2+\cdots+L_g$, and $1\leq
  L_i\leq L\leq M$.  For all integers $n_i\geq 1$ and all
  matrices $W_i\in \F_q^{L_i\times n_i}$, for $i\in[g]$,
  satisfying
  \[\sum_{i=1}^{g}\rank(W_i)\geq k,\]
  we have
  \begin{equation*}
  \begin{split}
  &\rank(D\diag(W_1,W_2,\dots,W_g))\\
  =&\rank((D(\gamma^0,\cB,k,L_1)W_1, D(\gamma^1,\cB,k,L_2)W_2,\\
   &\qquad\qquad\cdots, D(\gamma^{g-1},\cB,k,L_g)W_g))\\
  =&k.
  \end{split}
  \end{equation*}
  For the case
  \begin{equation*}
    \sum_{i=1}^{g}\rank(W_i)< k,
  \end{equation*}
  we have
  \begin{equation*}
    \begin{split}
      &\rank(D\diag(W_1,W_2,\dots,W_g))\\
      =&\rank((D(\gamma^0,\cB,k,L_1)W_1, D(\gamma^1,\cB,k,L_2)W_2,\\
      &\qquad\qquad \cdots, D(\gamma^{g-1},\cB,k,L_g)W_g))\\
      =&\sum_{i=1}^{g}\rank(W_i).
    \end{split}
  \end{equation*}
\end{theorem}
\begin{IEEEproof}
The first claim is exactly Lemma \ref{theorem_LRS}. For the second
one, we assume to the contrary that there { exist $W_i\in \F_q^{L_i\times n_i}$, for all $i\in[g]$,} with
\[\sum_{i=1}^{g}\rank(W_i)< k,\]
and
\begin{equation}
  \label{eq:cont}
  \rank(D\diag(W_1,W_2,\dots,W_g))<\sum_{i=1}^{g}\rank(W_i),
\end{equation}
where we apply a fact that $\rank(D\diag(W_1,W_2,\dots,W_g))\leq \rank(\diag(W_1,W_2,\dots,W_g))=\sum_{i=1}^{g}\rank(W_i)$.
Note that there { exist $W'_i\in \F_q^{L_i\times n'_i}$ for all $i\in[g]$,} such that
$\rank(W'_i)=n_i'$,
\[\sum_{i=1}^{g}\rank(W'_i)=k-\sum_{i=1}^{g}\rank(W_i),\]
and
\[\sum_{i=1}^{g}\rank(W_i,W_i')=k.\]
By the first claim,
\[\rank(D\diag((W_1,W'_1),(W_2,W'_2),\dots,(W_g,W'_g)))=k.\]
But now, combining this with ~\eqref{eq:cont}, we get
\begin{equation*}
\begin{split}
&\rank(D\diag(W'_1,W'_2,\dots,W'_g))\\
>&\sum_{i=1}^{g}n'_i
=\rank(\diag(W'_1,W'_2,\dots,W'_g)),
\end{split}
\end{equation*}
which is a contradiction. Thus, the desired result follows.
\end{IEEEproof}

\section{Code Construction}
\label{sec-construction-MRC}

In this section, we describe a construction for $(n,r,h,\delta,q)$-MR
codes. The main idea of our construction is to use generator matrices
of linearized Reed-Solomon codes for global parity-check symbols of MR
codes.

Throughout this section, we use the $(\delta-1)\times(r+\delta-1)$
matrix
  \begin{equation}\label{Eqn_P1}
    P_1\triangleq\begin{pmatrix}
    1&1&\cdots&1\\
    \alpha_1&\alpha_2&\cdots&\alpha_{r+\delta-1}\\
    \vdots&\vdots& &\vdots\\
    \alpha^{\delta-2}_1&\alpha^{\delta-2}_2&\cdots&\alpha^{\delta-2}_{r+\delta-1}\\
    \end{pmatrix}{ \in \F^{(\delta-1)\times(r+\delta-1)}_q},
  \end{equation}
  and the $h\times(r+\delta-1)$ matrix
  \begin{equation}\label{Eqn_P2}
    P_2\triangleq \begin{pmatrix}
      \alpha^{\delta-1}_{1}&\alpha^{\delta-1}_{2}&\dots &\alpha^{\delta-1}_{r+\delta-1}\\
      \alpha^{\delta}_1&\alpha_2^{\delta}&\dots &\alpha^{\delta}_{r+\delta-1}\\
      \vdots&\vdots& &\vdots\\
      \alpha^{\delta+h-2}_1&\alpha^{\delta+h-2}_2&\cdots &\alpha_{r+\delta-1}^{\delta+h-2}\\
    \end{pmatrix}{ \in \F^{h\times (r+\delta-1)}_q},
  \end{equation}
  where $\alpha_i\in \F_q\setminus\{0\}$, and $\alpha_i\ne \alpha_j$
  for $i\ne j$.  Let $\gamma_1,\gamma_2,\dots,\gamma_h\in\F_{q^h}$
  form a basis of $\F_{q^{h}}$ over $\F_q$. Define $\Gamma\triangleq
  (\gamma_1,\gamma_2,\dots,\gamma_h)\in \F^h_{q^h}$, and
  \begin{equation}\label{eqn_beta_coor}
   \bbeta \triangleq  (\beta_1,\beta_2,\dots,\beta_{r+\delta-1})=\Gamma P_2\in \F^{r+\delta-1}_{q^h},
  \end{equation}
  namely, each column of $P_2$ is translated to an element of
  $\F_{q^h}$.

\begin{construction}\label{cons_PMDS}   For $m\in \N$, let $\cC$ be the linear code
{ with length $n$ over $\F_{q^{h}}$}  given by the
  parity-check matrix
 \begin{equation}\label{eqn_def_H}
 \begin{split}
    &H\triangleq\\
    &\begin{pmatrix}
    P_1&0&\cdots&0\\
    0&P_1&\cdots&0\\
    \vdots&\vdots&\ddots&\vdots\\
    0&0&\cdots&P_1\\
    D(\gamma^0,\bbeta,h,a)&D(\gamma^1,\bbeta,h,a)&\cdots&D(\gamma^{m-1},\bbeta,h,a)
    \end{pmatrix},
    \end{split}
  \end{equation}
  where $\gamma\in \F_{q^h}$ is a primitive element and $a=r+\delta-1$.
\end{construction}

\begin{theorem}\label{theorem_main}
  Let $q\geq \max\{r+\delta,m+1\}$.  Then the code
  $\cC$ from Construction~\ref{cons_PMDS} is an
  $(n=m(r+\delta-1),r,h,\delta,q^h)$-MR code with the minimum Hamming distance
  $d=(\lfloor\frac{h}{r}\rfloor+1)(\delta-1)+h+1$.
\end{theorem}
\begin{IEEEproof}
  To simplify the notation, let us denote the
  $((i-1)(r+\delta-1)+j)$th coordinate by the pair $(i,j)$, where
  $i\in[m]$ and $j\in[r+\delta-1]$. Using this notation, the $i$th
  repair set is given by $S_i=\{(i,j)~:~j\in[r+\delta-1]\}$, for
  $i\in[m]$.

  Recall from \eqref{Eqn_P1} that $P_1$ is a Vandermonde matrix.
  Therefore, by~\eqref{eqn_def_H}, { $\cC|_{S_i}$ is a subcode of an
  $[r+\delta-1,r,\delta]_q$ MDS code}, which implies that the code
  $\cC$ has $(r,\delta)_a$-locality. We shall now prove the code can
  recover from all erasure patterns
  $\cE=\{E_{i_1},E_{i_2},\dots,E_{i_t}\}$ such that $E_{i_\ell}\subseteq
  S_{i_\ell}$, $|E_{i_\ell}|\geq \delta$, and
  \begin{equation}\label{eqn_res_dim}
    \sum_{\ell=1}^{t}|E_{i_\ell}|-t(\delta-1)\leq h,
  \end{equation}
  namely, $\cC$ is an $(n,r,h,\delta,q^h)$-MR code.

  For $\ell\in[t]$, assume
  $E_{i_\ell}=\{(i_\ell,j_1),(i_\ell,j_2),\dots,(i_\ell,j_{|E_{i_\ell}|})\}$,
  and the columns of $P_1$ are denoted by
  $P_1=(P_{1,1},P_{1,2},\dots,P_{1,r+\delta-1})$.  Define the
  projections of $P_1$ and $D(\gamma^{i-1},\bbeta,h,r+\delta-1)$ onto
  the erased coordinates as
  \[P_1|_{E_{i_\ell}}\triangleq(P_{1,j_1},P_{1,j_2},\cdots, P_{1,j_{|E_{i_\ell}|}}),\]
  and
  \begin{equation}\label{Eqn_DE}
  \begin{split}
    &D(\gamma^{i-1},\bbeta,h,r+\delta-1)|_{E_{i_\ell}}\\
    \triangleq& \parenv{
      \begin{matrix}
        \beta_{j_1} &\beta_{j_2}&\cdots&\beta_{j_{|E_{i_\ell}|}}\\
        \cD^1_{\gamma^{i-1}}(\beta_{j_1}) &\cD^1_{\gamma^{i-1}}(\beta_{j_2})& \cdots &\cD^1_{\gamma^{i-1}}(\beta_{j_{|E_{i_\ell}|}})\\
        \vdots &\vdots& &\vdots\\
        \cD^{h-1}_{\gamma^{i-1}}(\beta_{j_1}) &\cD^{h-1}_{\gamma^{i-1}}(\beta_{j_2})& \cdots &\cD^{h-1}_{\gamma^{i-1}}(\beta_{j_{|E_{i_\ell}|}})\\
    \end{matrix}}.
    \end{split}
  \end{equation}
  Proving that $\cE$ is recoverable is equivalent to showing that the matrix
  \begin{equation*}
    H_{\cE}\triangleq \begin{pmatrix}
      P_1|_{E_{i_1}}&0&\cdots&0\\
      0&P_1|_{E_{i_2}}&\cdots&0\\
      \vdots&\vdots&\ddots&\vdots\\
      0&0&\cdots&P_1|_{E_{i_t}}\\
      D_{i_1,E_{i_1}}&D_{i_2,E_{i_2}}&\cdots&D_{i_t,E_{i_t}}
    \end{pmatrix}
  \end{equation*}
  has full column rank, where $D_{i_\ell,E_{i_\ell}}=D(\gamma^{i_\ell-1},\bbeta,h,a)|_{E_{i_\ell}}$ for $\ell\in[t]$.
  Otherwise, we cannot distinguish between a codeword $C\in \cC$
  from $C+C'$, where the nonzero components of $C'$ is a nonzero solution of $H_{\cE}X=0$.

  Since $P_1$ is a Vandermonde matrix, for any $E^*_{i_\ell}\subseteq
  E_{i_\ell}$ with $|E^*_{i_\ell}|=\delta-1$, $\ell\in[t]$, we have
  that $P_1|_{E^*_{i_\ell}}$ has full rank. Denote
  $\overline{E}_{i_\ell}=E_{i_\ell}\setminus E^*_{i_\ell}$. Thus,
  there exists a matrix $A_{i_\ell}\in
  \F^{|E^*_{i_\ell}|\times|\overline{E}_{i_\ell}|}_q$ such that
  \begin{equation}\label{eqn_P_W_j}
    \begin{pmatrix}
      P_1|_{E^*_{i_\ell}} &P_1|_{\overline{E}_{i_\ell}}\\
      P_2|_{E^*_{i_\ell}} &P_2|_{\overline{E}_{i_\ell}}\\
    \end{pmatrix}
    \begin{pmatrix}
      I_{E^*_{i_\ell}} &-A_{i_\ell}\\
      0& I_{\overline{E}^*_{i_\ell}}
    \end{pmatrix}=\begin{pmatrix}
    P_1|_{E^*_{i_\ell}} &0\\
    P_2|_{E^*_{i_\ell}}& W_{i_\ell}
    \end{pmatrix}
  \end{equation}
  i.e.,
  \begin{equation}\label{Eqn_P1E}
   P_1|_{\overline{E}_{i_\ell}}=P_1|_{E^*_{i_\ell}} A_{i_\ell},
  \end{equation}
  and
  \begin{equation}\label{Eqn_WP2}
    W_{i_\ell}=P_2|_{\overline{E}_{i_\ell}}-(P_2|_{E^*_{i_\ell}})A_{i_\ell},
  \end{equation}
  where $W_{i_\ell}$ is an $h\times |\overline{E}_{i_\ell}|$ matrix
  over $\F_q$.  Denote
  \begin{equation}\label{eqn_beta_star}
    \bbeta_{i_\ell}^*=\Gamma W_{i_\ell}
  \end{equation}
  for $\ell\in[t]$.    For $\tau\in \overline{E}_{i_\ell}$,  write
  \begin{equation}\label{Eqn_P1W}
    P_{1,\tau}=\sum_{a\in E^*_{i_\ell}}e^{(i_\ell)}_{a,\tau}P_{1,a}
  \end{equation}
  with $e^{(i_\ell)}_{a,\tau}\in \F_q$ determined by
  $A_{i_\ell}$. Then, it follows from \eqref{eqn_beta_coor} and
  \eqref{Eqn_P1E}-\eqref{Eqn_P1W} that
  \begin{equation*}
    \beta^*_{i_\ell,\tau}=\beta_{\tau}-
    \sum_{a\in E^*_{i_\ell}}e^{(i_\ell)}_{a,\tau}\beta_{a}.
  \end{equation*}

  {
 Note that
  \begin{equation*}
  \begin{split}
  &D(\gamma^{i_{\ell}-1},\bbeta,h,r+\delta-1)|_{E_{i_\ell}}\begin{pmatrix}
      I_{E^*_{i_\ell}} &-A_{i_\ell}\\
      0& I_{\overline{E}^*_{i_\ell}}
    \end{pmatrix}\\
    =&D(\gamma^{i_\ell-1},\Gamma(P_2|_{E^*_{i_\ell}},P_2|_{\overline{E}_{i_\ell}}),h,|E_{i_\ell}|)\begin{pmatrix}
      I_{E^*_{i_\ell}} &-A_{i_\ell}\\
      0& I_{\overline{E}^*_{i_\ell}}
    \end{pmatrix}\\
    =&D(\gamma^{i_\ell-1},\Gamma(P_2|_{E^*_{i_\ell}}, W_{i_\ell}),h,|E_{i_\ell}|)\\
    =&(D_{i_\ell},D(\gamma^{i_\ell-1},\bbeta_{i_\ell}^*,h,|\overline{E}_{i_\ell}|)),\\
    \end{split}
    \end{equation*}
    where the second equality holds by the linearity of $\cD^{i}_{\alpha}(\cdot)$
    (Proposition \ref{prop_linearity}) and~\eqref{eqn_P_W_j}, and
    the last equality holds by~\eqref{eqn_beta_star}.
  This is to say that $H_{\cE}$ is equivalent with
  \[
   \begin{pmatrix}
      P_1|_{E^*_{i_1}}&0&0&0&\cdots&0&0\\
      0&0&P_1|_{E^*_{i_2}}&0&\cdots&0&0\\
      \vdots&\vdots&\vdots&\vdots& &\vdots&\vdots\\
      0&0&0&0&\cdots&P_1|_{E^*_{i_t}}&0\\
      D_{i_1}&D^*_{i_1,\overline{E}_{i_1}}&
      D_{i_2}&D^*_{i_2,\overline{E}_{i_2}}&\cdots&
      D_{i_t}&D^*_{i_t,\overline{E}_{i_t}}
    \end{pmatrix},
  \]
  where $D^*_{i_\ell,\overline{E}_{i_\ell}}=D(\gamma^{i_\ell-1},\bbeta_{i_\ell}^*,h,|\overline{E}_{i_\ell}|)$
  for $\ell\in[t]$.
  Recall that $P_1|_{E^*_{i_j}}$ for $j\in[t]$ has full rank.
  Hence, $H_{\cE}$ is equivalent with
  \begin{equation*}
  \begin{split}
    &H^*_{\cE}\triangleq\\
    &\begin{pmatrix}
      P_1|_{E^*_{i_1}}&0&0&0&\cdots&0&0\\
      0&0&P_1|_{E^*_{i_2}}&0&\cdots&0&0\\
      \vdots&\vdots&\vdots&\vdots& &\vdots&\vdots\\
      0&0&0&0&\cdots&P_1|_{E^*_{i_t}}&0\\
      0&D^*_{i_1,\overline{E}_{i_1}}&
      0&D^*_{i_2,\overline{E}_{i_2}}&\cdots&
      0&D^*_{i_t,\overline{E}_{i_t}}
    \end{pmatrix},
    \end{split}
  \end{equation*}}
  where $D^*_{i_\ell,\overline{E}_{i_\ell}}=D(\gamma^{i_\ell-1},\bbeta_{i_\ell}^*,h,|\overline{E}_{i_\ell}|)$
  for $\ell\in[t]$.
 Then, $H_{\cE}$ has full column rank if
  and only if
  \begin{equation*}
  \begin{split}
  &(D^*_{i_1,\overline{E}_{i_1}},D^*_{i_1,\overline{E}_{i_1}},\cdots,D^*_{i_1,\overline{E}_{i_1}})\\
  =&(D(\gamma^{i_1-1},\bbeta_{i_1}^*,h,|\overline{E}_{i_1}|),
  D(\gamma^{i_2-1},\bbeta_{i_2}^*,h,|\overline{E}_{i_2}|),\\
  &\qquad\cdots,
  D(\gamma^{i_t-1},\bbeta_{i_t}^*,h,|\overline{E}_{i_t}|))
  \end{split}
  \end{equation*}
   has full
  column rank. Note from \eqref{Eqn_P1} and \eqref{Eqn_P2}, that
  $\parenv{\begin{smallmatrix} P_1\\ P_2 \end{smallmatrix}}$ forms an
  $(h+\delta-1)\times (r+\delta-1)$ Vandermonde matrix. Clearly,
  $|E_{i_\ell}|\leq \min\{h+\delta-1,r+\delta-1\}$ for $\ell\in[t]$,
  which means
  \begin{equation*}
    \rank\begin{pmatrix}
        P_1|_{E^*_{i_\ell}} &P_1|_{\overline{E}_{i_\ell}}\\
        P_2|_{E^*_{i_\ell}} &P_2|_{\overline{E}_{i_\ell}}\\
      \end{pmatrix}
    =|E^*_{i_\ell}|+|\overline{E}_{i_\ell}|,
  \end{equation*}
  and $\rank(P_1|_{E^*_{i_\ell}})=|E^*_{i_\ell}|.$ Thus,
  \eqref{eqn_P_W_j} implies
  $\rank(W_{i_\ell})=|\overline{E}_{i_\ell}|$ for $\ell\in[t]$.
  Now, according to \eqref{Eqn_Def_D}, \eqref{eqn_beta_star} and the linearity  of $\cD^i_\alpha(\cdot)$, we have
  \begin{eqnarray}
      &&\rank((D(\gamma^{i_1-1},\bbeta_{i_1}^*,h,|\overline{E}_{i_1}|),
      D(\gamma^{i_2-1},\bbeta_{i_2}^*,h,|\overline{E}_{i_2}|),\notag\\
      &&\qquad\cdots,
      D(\gamma^{i_t-1},\bbeta_{i_t}^*,h,|\overline{E}_{i_t}|)))\notag\\
      &=&\rank((D(\gamma^{i_1-1},\Gamma,h,h)W_{i_1},
      D(\gamma^{i_2-1},\Gamma,h,h)W_{i_2},\notag\\
      &&\qquad\cdots,
      D(\gamma^{i_t-1},\Gamma,h,h)W_{i_t}))\notag\\
      &=&\rank((D(\gamma^{0},\Gamma,h,h)W'_1,
      D(\gamma^{1},\Gamma,h,h)W'_2,\notag\\
      &&\qquad\cdots,
      D(\gamma^{m-1},\Gamma,h,h)W'_m)),\label{Eqn_DW}
  \end{eqnarray}
  where
  \begin{equation}\label{Eqn_W'}
  W'_i\triangleq\begin{cases}
  W_i,& \text{if }i\in \{i_\ell~:~\ell\in [t]\},\\
  0,& \text{otherwise.}
  \end{cases}
  \end{equation}
  We observe that
  \begin{eqnarray*}
  (D(\gamma^{0},\Gamma,h,h),
  D(\gamma^{1},\Gamma,h,h),\cdots, D(\gamma^{m-1},\Gamma,h,h))
  \end{eqnarray*}
  can be regarded as the generator matrix of a linearized Reed-Solomon
  code with parameters $[mh,h]_{q^h}$ according to
  Definition~\ref{def_LRS}.
Then, applying Theorem~\ref{thm_rank} to \eqref{Eqn_DW} and \eqref{Eqn_W'},
  we conclude that
  \begin{equation*}
    \begin{split}
      &\rank((D(\gamma^{i_1-1},\bbeta_{i_1}^*,h,|\overline{E}_{i_1}|),
      D(\gamma^{i_2-1},\bbeta_{i_2}^*,h,|\overline{E}_{i_2}|),\\
      &\qquad\cdots,
      D(\gamma^{i_t-1},\bbeta_{i_t}^*,h,|\overline{E}_{i_t}|)))\\
      =&\sum_{i=1}^{m}\rank(W'_{i})\\
      =&\sum_{\ell=1}^{t}\rank(W_{i_\ell})\\
      =&\sum_{\ell=1}^{t}|\overline{E}_{i_\ell}|,
    \end{split}
  \end{equation*}
  which means $H^*_{\cE}$ has full rank, i.e., $H_{\cE}$ has
  full rank for all possible $\cE$ that satisfy~\eqref{eqn_res_dim}.
  Therefore,  $\cC$ can recover all the erasure patterns required by MR
  codes.

  Having reached this point, the desired result follows from the fact that
  MR codes are optimal LRCs. Hereafter, for the sake of completeness, we derive the minimum Hamming
  distance for the reader's convenience. We know the code $\cC$ can recover from
  any erasure pattern that affects at most $\delta-1$ coordinates in
  each repair set, and any additional $h$ erased positions. Let us
  consider the other erasure patterns, obviously where all the
  affected repair sets have at least $\delta$ erasures each. In
  particular, we consider the \emph{minimal} erasure configurations,
  namely, configurations in which the removal of any one erasure makes
  it recoverable. Assume that $a$ repair sets are affected.  Then, the
  total number of erasures is $a(\delta-1)+h+1$, where the $h+1$
  erasures are distributed among the $a$ affected repair sets, i.e.,
  it requires $a(\delta-1)+h+1\leq a(r+\delta-1)$ and thus
  \[a\geq \ceilenv{\frac{h+1}{r}}=\floorenv{\frac{h}{r}}+1.\]
  Therefore, a lower bound on the Hamming distance of $\cC$ is
  \[ d\geq \parenv{\floorenv{\frac{h}{r}}+1}(\delta-1)+h+1.\]

  Note from \eqref{eqn_def_H} that  $k\geq n-h-m(\delta-1)=mr-h$ which
  implies $\ceilenv{\frac{k}{r}}+\floorenv{\frac{h}{r}}\geq m$.
  Since $\cC$ is a locally repairable code with
  $(r,\delta)_a$-locality, by Lemma~\ref{lemma_bound_i} we have
  \begin{align*}
      d&\leq n-k-\parenv{\ceilenv{\frac{k}{r}}-1}(\delta-1)+1\\
      &\leq n-k-\parenv{m-\floorenv{\frac{h}{r}}-1}(\delta-1)+1\\
      &\leq h+\parenv{\floorenv{\frac{h}{r}}+1}(\delta-1)+1.
  \end{align*}
  Combining this with the lower bound on $d$, we obtain
  \[d=\parenv{\floorenv{\frac{h}{r}}+1}(\delta-1)+h+1.\]
  Thus, $\cC$ is an $(n,r,h,\delta,q^h)$-MR code with $d=(\lfloor\frac{h}{r}\rfloor+1)(\delta-1)+h+1$.
\end{IEEEproof}

\begin{corollary}
  \label{cor:opt}
  Let $q\geq \max\{r+\delta,m+1\}$ and $\delta\geq 2$. If
  $m=\Theta(q)$ and $r=\Theta(q)$ (implying $n=\Theta(q^2)$), then for
  fixed $h\leq \min\{m,\delta+1\}$ the code $\cC$ generated by
  Construction~\ref{cons_PMDS} is an
  $(n=m(r+\delta-1),r,h,\delta,q^h)$-MR code with asymptotically
  order-optimal field size $q^{h}=\Theta(n^{h/2})$.
\end{corollary}
\begin{IEEEproof}
  By our setting, the field size of the code generated by
  Construction~\ref{cons_PMDS} is $\Theta(q^{h})$. According
  to Lemma~\ref{lemma_lower_bound_F}, the field size { must be} at least
  $$\Omega(nr^{\min\{\delta-1,h-2\}})=\Omega(m(r+\delta-1)r^{h-2})=\Omega(q^{h}),$$
  where the first equality holds by $h\leq \delta+1$, and the second
  one follows from $m=\Theta(q)$, $r=\Theta(q)$, and the fact that
  $h$, $\delta$ are regarded as constants.  Thus, the code $\cC$
  generated by Construction~\ref{cons_PMDS} has asymptotically
  order-optimal field size $\Theta(q^{h})$.
\end{IEEEproof}

\begin{example}

Let $r=2$, $\delta=2$, $q=4$, and $m=3$. By Construction \ref{cons_PMDS} and Theorem \ref{theorem_main},
an $(n=9,r=2,h=2,\delta=2,q^2=16)$-MR code can be given by the following parity-check matrix
\begin{equation*}
\begin{pmatrix}
1&1&1&0&0&0&0&0&0\\
0&0&0&1&1&1&0&0&0\\
0&0&0&0&0&0&1&1&1\\
\alpha^{6}& \alpha^{9}& \alpha^{10}& \alpha^{7}& \alpha^{10}& \alpha^{11}& \alpha^{8}& \alpha^{11}& \alpha^{12}\\
\alpha^{10}& \alpha^{7}& \alpha^{11}& \alpha^{0}& \alpha^{12}& \alpha^{1}& \alpha^{5}& \alpha^{2}& \alpha^{6}\\
\end{pmatrix},
\end{equation*}
where $\alpha$ is a primitive element of $\F_{16}$.
\end{example}

\section{Concluding Remarks}\label{sec-conclusion}

In this paper, we introduced a construction of maximally recoverable
codes with uniform-sized disjoint repair sets, also known as partial
MDS (PMDS) codes. Our construction is based on linearized Reed-Solomon
codes, and it yields maximally recoverable codes with field size
$\Theta((\max\{r+\delta-1,\frac{n}{r+\delta-1}\})^{h})$, where $h$ and
$\delta$ are constants.  Compared with known constructions, our
construction can generate maximally recoverable codes with a smaller
field size in certain cases. In some particular regimes, described in
Corollary~\ref{cor:opt}, the construction produces code families with
order-optimal field size. For more details about parameters for MR
codes, a summary of the results in comparison with known constructions
is given in Table~\ref{tab:comp}, where $q$ and $q_1$ are prime
powers, and $m=\frac{n}{r+\delta-1}$.

We would like to highlight some interesting cases from
Table~\ref{tab:comp}. In \cite{gopi2020maximally}, a construction for
$(n,r,3,\delta,q)$-MR codes was provided, achieving $q=\Theta(n^3)$,
but only for odd characteristic. Finding a comparable construction for
even characteristic was left as an open question. Here,
Construction~\ref{cons_PMDS} provides an answer to this question,
since our construction does not impose a restriction on the parity of
the field characteristic, and it achieves the same order
$q=\Theta(n^3)$.

Another case we would like to point out involves the asymptotic regime
where $r=\Theta(n)$. In this regime, our construction achieves a field
size of $q=\Theta(n^h)$. For odd $q$ or $\delta>2$, this improves upon
the best known construction from~\cite{gabrys2018constructions}, which
achieves $q=\Theta(n^{h\delta})$. When $\delta=2$, $q$ is even, and
$r=\Theta(n)$, the best known result is still the one in
\cite{gopalan2014explicit} with $q=\Theta(k^{h-1})=\Theta(n^{h-1})$.


  In addition, \cite{gabrys2018constructions} challenged
  researchers to find families of PMDS codes with smaller field sizes
  than $\max\{m,(r+\delta-1)^{h+\delta-1}\}^{h}$.  The construction
  in~\cite{martinez2019universal} does so for the case $h<r$ and
  $(r+\delta-1)^{h+\delta-1}>m$. Similarly, the construction
  in~\cite{bogart2020constructing} also improves
  upon~\cite{gabrys2018constructions} for the case $r=2$. In this
  paper, the MR codes generated by Construction \ref{cons_PMDS}
  provide an improvement over~\cite{gabrys2018constructions} for
  $(r+\delta-1)^{h+\delta-1}>m$, since in this case
  $\max\{r+\delta-1,\frac{n}{r+\delta-1}\}^{h}<\max\{m,(r+\delta-1)^{h+\delta-1}\}^{h}$.

The broad problem of closing the gap between the field-size
requirements of known constructions and the theoretic bounds is still
largely open. Further closing this gap, beyond the results of this
paper, is left for future work.

\section*{Acknowledgments}

The authors would like to thank the Associate Editor, Prof.~Camilla Hollanti and the anonymous reviewers, whose comments and suggestions
improved the presentation of this paper.

\bibliographystyle{IEEEtranS}

\bibliography{HanBib}

\begin{IEEEbiographynophoto}{Han Cai}(S'16-M'18)
received the B.S. and M.S. degrees in mathematics from Hubei
University, Wuhan, China, in 2009 and 2013, respectively and
received the Ph.D.
degree from the Department of Communication
Engineering, Southwest Jiaotong
University, Chengdu, China, in 2017.
During Oct. 2015 to Oct. 2017, he was a visiting
Ph.D. student in the Faculty of Engineering, Information and Systems, University of Tsukuba, Japan.
From 2018 to 2021, he was a postdoctoral fellow at the School of
Electrical \& Computer Engineering, Ben-Gurion University of the Negev, Israel.
In 2021, he joined Southwest Jiaotong
University, where he currently hold a tenure-track
position.
His research interests
include coding theory and sequence design.
\end{IEEEbiographynophoto}

\begin{IEEEbiographynophoto}{Ying Miao}
received the D.Sci. degree in mathematics from Hiroshima University, Hiroshima, Japan, in 1997.

From 1989 to 1993, he worked for Suzhou Institute of Silk Textile Technology, Suzhou,
Jiangsu, P. R. China. From 1995 to 1997, he was a Research Fellow
of the Japan Society for the Promotion of Science.
During 1997--1998, he was a Postdoctoral Fellow at the Department of Computer Science,
Concordia University, Montreal, QC, Canada.
In 1998, he joined the University of Tsukuba, Tsukuba, Ibaraki, Japan,
where he is currently a Full Professor
at the Faculty of Engineering, Information and Systems.
His current research interests include combinatorics, coding theory, and information security.

Dr. Miao is on the Editorial Boards of several journals such as Graphs and Combinatorics, and Journal of Combinatorial Designs.
He received the 2001 Kirkman Medal from the Institute of Combinatorics and its Applications.
\end{IEEEbiographynophoto}

\begin{IEEEbiographynophoto}{Moshe Schwartz}
(Senior Member, IEEE)
is a professor in the School of Electrical and Computer
Engineering, Ben-Gurion University of the Negev, Israel. His research
interests include algebraic coding, combinatorial structures, and
digital sequences.

Prof.~Schwartz received the B.A.~(\emph{summa cum laude}), M.Sc., and
Ph.D.~degrees from the Technion -- Israel Institute of Technology,
Haifa, Israel, in 1997, 1998, and 2004 respectively, all from the
Computer Science Department. He was a Fulbright post-doctoral
researcher in the Department of Electrical and Computer Engineering,
University of California San Diego, and a post-doctoral researcher in
the Department of Electrical Engineering, California Institute of
Technology. While on sabbatical 2012--2014, he was a visiting scientist
at the Massachusetts Institute of Technology (MIT).

Prof.~Schwartz received the 2009 IEEE Communications Society Best
Paper Award in Signal Processing and Coding for Data Storage, and the
2020 NVMW Persistent Impact Prize. He served as an Associate Editor
for Coding Techniques and coding theory for the IEEE Transactions on
Information Theory during 2014--2021, and since 2021 he has been
serving as an Area Editor for Coding and Decoding for the IEEE
Transactions on Information Theory.  He is also an Editorial Board
Member for the Journal of Combinatorial Theory Series A since 2021.
\end{IEEEbiographynophoto}

\begin{IEEEbiographynophoto}{Xiaohu Tang}
 (M'04-SM'18)  received the B.S. degree in applied mathematics from
the Northwest Polytechnic University, Xi'an, China, the M.S. degree in applied
mathematics from the Sichuan University, Chengdu, China, and the Ph.D.
degree in electronic engineering from the Southwest Jiaotong University,
Chengdu, China, in 1992, 1995, and 2001 respectively.

From 2003 to 2004, he was a research associate in the Department of Electrical
and Electronic Engineering, Hong Kong University of Science and Technology.
From 2007 to 2008, he was a visiting professor at University of Ulm,
Germany. Since 2001, he has been in the School of Information Science and Technology,
Southwest Jiaotong University, where he is currently a professor. His research
interests include coding theory, network security, distributed storage and information processing for big data.

Dr. Tang was the recipient of the National excellent Doctoral Dissertation
award in 2003 (China), the Humboldt Research Fellowship in 2007
(Germany), and the Outstanding Young Scientist Award by NSFC in 2013
(China). He served as Associate Editors for several journals including \textit{IEEE
Transactions on Information Theory} and \textit{IEICE Transactions on
Fundamentals}, and served on a number of technical program committees of
conferences.
\end{IEEEbiographynophoto}

\end{document}